\documentclass[aps,prb,reprint,amsmath,amssymb,floatfix,superscriptaddress]{revtex4-2}

\usepackage{graphicx}
\usepackage{overpic}
\usepackage{gensymb}
\usepackage{siunitx}

\usepackage{xcolor}

\usepackage{soul}

\makeatletter

\newcommand{\appsection}[1]{\section{\uppercase{#1}}}
\makeatother

\begin{document}

\title{Out-of-plane interference effects in the angle-resolved photoemission of epitaxial graphene}

\author{E. Favorite}
\email{ejfavori@uark.edu}
\affiliation{Department of Physics, University of Arkansas, Fayetteville, AR 72701, USA}
\author{P. Krüger}
\affiliation{Graduate School of Engineering and Molecular Chirality Research Center, Chiba University, Chiba 263-8522, Japan}
\author{Ph. Rosenzweig}
\affiliation{Physikalisches Institut, Universität Stuttgart, Pfaffenwaldring 57, 70569 Stuttgart, Germany}
\author{R. Joshi}
\affiliation{Department of Physics, University of Arkansas, Fayetteville, AR 72701, USA}
\author{S. Williams}
\affiliation{Department of Physics, University of Arkansas, Fayetteville, AR 72701, USA}
\author{B. Matta}
\affiliation{Max-Planck-Institut für Festkörperforschung, Heisenbergstraße 1, 70569 Stuttgart, Germany}
\author{K. Küster}
\affiliation{Max-Planck-Institut für Festkörperforschung, Heisenbergstraße 1, 70569 Stuttgart, Germany}
\author{M. Hashimoto}
\affiliation{Stanford Synchrotron Radiation Lightsource, SLAC National Accelerator Lab, USA}
\author{U. Starke}
\affiliation{Max-Planck-Institut für Festkörperforschung, Heisenbergstraße 1, 70569 Stuttgart, Germany}
\author{H. Nakamura}
\email{hnakamur@uark.edu}
\affiliation{Department of Physics, University of Arkansas, Fayetteville, AR 72701, USA}

%Abstract
\begin{abstract}
The epitaxial growth of monolayer graphene by Si sublimation of 6H-SiC(0001) is one of the most effective ways of producing large-scale monocrystalline graphene samples of high quality. However, graphene is left interacting with a reconstructed carbon buffer layer known as zerolayer graphene (ZLG) and the SiC substrate underneath. Beyond the mere creation of Dirac cone replicas, the effect of this interaction on photoemission measurements is poorly understood. We utilized synchrotron radiation for photon-energy-dependent angle-resolved photoemission spectroscopy (ARPES) to investigate the complex final-state interactions of photoelectrons within the heterostructure of monolayer graphene, ZLG, and SiC. We identify multiple resonances in the photoemission intensity of the Dirac cone as a function of photon energy that originate from interlayer interference, confirmed with a multiple scattering model. We also demonstrate for the first time that the relative intensity of the replica bands induced by the $(6\sqrt{3} \times 6\sqrt{3})\text{R}30^\circ$ reconstruction scales with the electron kinetic energy. Finally, we resolved four nondispersing electronic states, which originate from the ZLG. Our findings show that the three-dimensional (3D) structure of SiC and ZLG manifests itself in the photoemission from two-dimensional (2D) epitaxial graphene. By revealing how 3D substrate architecture dictates the photoemission of a 2D layer, this work establishes a crucial framework for investigating the electronic structure of complex heterostructures via ARPES.
\end{abstract}

\maketitle

\section{Introduction}
The emergence of two-dimensional (2D) materials, especially graphene, has revolutionized the field of condensed matter physics and opened new possibilities for tunable materials by stacking to form heterostructures \cite{novoselov2012roadmap, li2024towards}. Arguably the best platform for producing large-scale, high-quality graphene samples is by thermal decomposition of silicon carbide (SiC) \cite{emtsev2009towards, kusunoki2015growth}. This method can produce monolayer graphene (MLG) at a wafer scale. However, underneath the MLG lies another, $(6\sqrt{3} \times 6\sqrt{3})\text{R}30^\circ$ reconstructed, carbon layer. This layer has the typical honeycomb structure of graphene, but is in part covalently bonded to the SiC. Therefore it does not possess graphene's typical $\pi$-band structure and is called zerolayer graphene (ZLG). The ZLG can interact with and strain the MLG \cite{schumann2014effect}. Together the MLG, ZLG, and SiC form a complex three-dimensional (3D) heterostructure. To understand this system, it is important to measure the electron energy dispersion.

In angle-resolved photoelectron spectroscopy (ARPES), understanding the matrix element effect on measurements is of the utmost importance to effectively study a material \cite{damascelli_probing_2004}. It is well understood how the matrix element depends on properties of the material such as orbital symmetry, atomic locations, and material thickness as well as the choice of photon energy $h\nu$, polarization, and angle of incidence to the sample \cite{moser_experimentalists_2017, bansil1999importance, ohta_interlayer_2007}. The best known example of this effect in graphene is the so-called \textit{dark corridor} effect, in which pseudospin suppresses the photoemission intensity in parts of the Fermi surface \cite{ shirley_brillouin-zone-selection_1995, daimon1995structure}. Work was done to leverage the matrix element by varying polarization to allow for the full visibility of the Fermi surface \cite{gierz_illuminating_2011, ayria_photon_2015}. Furthermore, recent studies have demonstrated that $h\nu$ tuning is critical for modulating the intensity distribution around the $\overline{\mathrm{K}}$ and $\overline{\Gamma}$ points, providing a pathway to investigate the final-state effects in 2D systems \cite{kern_simple_2023, nozaki_computational_2024, ayria_photon_2015}. However, while these models effectively capture the intrinsic symmetry of the graphene lattice, they typically ignore how the presence of the substrate will affect the photoemission process. A significant knowledge gap remains in how the 3D structure of the MLG/ZLG/SiC system fundamentally changes the matrix element of the final-state electron. Additionally, previous $h\nu$-dependent studies \cite{kern_simple_2023, nozaki_computational_2024, ayria_photon_2015} have generally been restricted to limited energy ranges, which may fail to resolve high-energy resonances or the full evolution of many-body features.

In this paper, we address these gaps by performing systematic ARPES measurements over a wide photon energy range of \SIrange{20}{200}{\electronvolt} allowing a better understanding of 3D substrate-driven final-state effects. By combining ARPES results with multiple scattering (MS) simulations we aim to isolate the intrinsic matrix element effects of MLG from 3D scattering. First, we establish photoelectron intensity resonances due to constructive interferences within the substrate scattering path and relate them to out-of-plane layer distances. Second, we investigate the dependence of the $(6\sqrt{3} \times 6\sqrt{3})\text{R}30^\circ$ replica band intensity on $h\nu$ and demonstrate that a simple explanation related to the inelastic mean free path (IMFP) is insufficient. Third, we identify the optimal photon energy for resolving electron-phonon coupling in MLG. Finally, we report four interfacial states arising from the bonding of the ZLG to SiC. Our findings reinforce that an understanding of final-state effects is a prerequisite for the accurate determination of electronic properties of supported 2D materials.

\section{Experiment and Numerical Methods} \label{methods}

\subsection{Sample preparation}
\begin{figure}[htbp]
    \centering
    \includegraphics[width=1\linewidth]{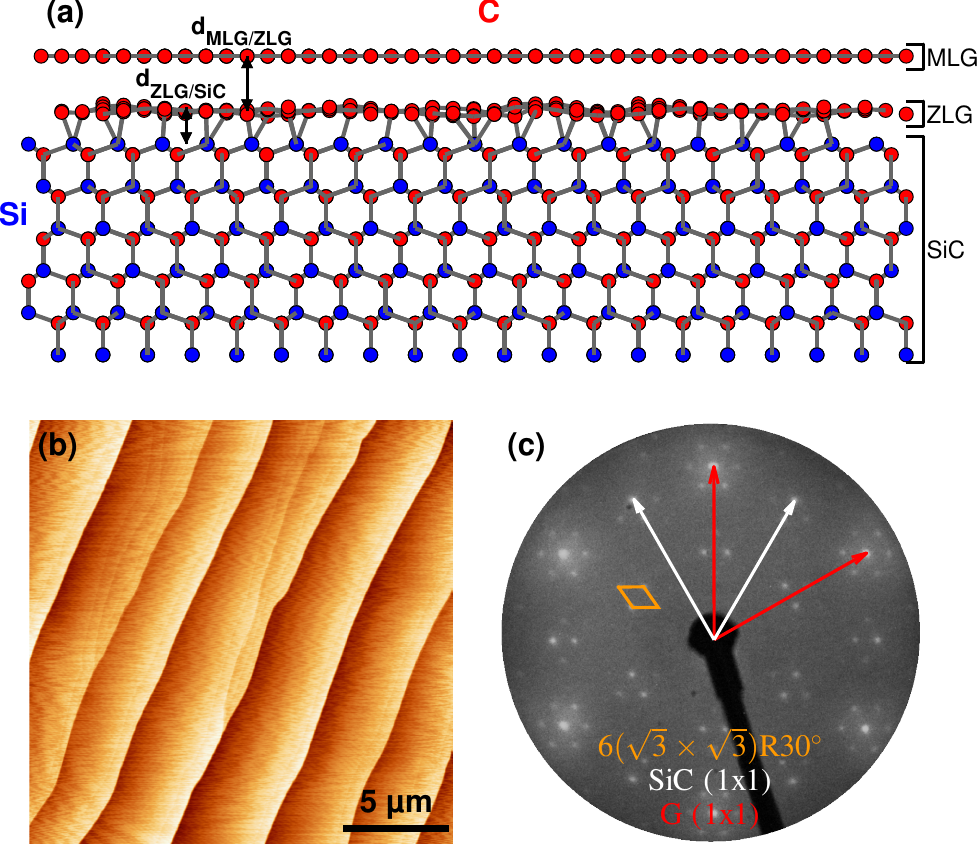}
    \caption{(a) Schematic ball-and-stick model (side view) of the sample crystal structure showing the MLG and ZLG layers and the SiC substrate. The ZLG exhibits vertical buckling  due to partial covalently bonding to the substrate and the MLG is interacting with the ZLG via van der Waals forces. The nominal layer separation distances are set to $d_{\text{SiC-ZLG}} = \SI{2.00}{\angstrom}$ and $d_{\text{ZLG-MLG}} = \SI{3.24}{\angstrom}$ \cite{razado2018structural}. (b) AFM image of a $20\times \qty{20}{\square\micro\metre}$ region of the sample showing uniform terraces. (c) LEED image taken at \SI{80}{\electronvolt} with reciprocal lattice vectors for graphene (red), SiC (white) indicated and a diamond highlighting typical spots of the $(6\sqrt{3} \times 6\sqrt{3})\text{R}30^\circ$ reconstructed ZLG (orange).}
    \label{fig:Sample}
\end{figure}

Single-crystalline 6H-SiC(0001) wafer pieces were used for the epitaxial growth of MLG. The substrate was first cleaned using a wet etching process. The sample was then placed in an induction heated graphite susceptor where it was brought to a temperature of \SI{1400}{\degreeCelsius} under a flow of hydrogen for 20 min. This process results in a clean substrate of unit-cell-high steps with terrace widths on the order of \SI{1}{\micro\metre} \cite{soubatch2005structure}. The substrate was then heated to \SI{1420}{\degreeCelsius} for 15 min under a flow of argon to induce graphitization via Si sublimation up to MLG coverage \cite{emtsev2009towards}. Figure~\ref{fig:Sample}(a) shows a schematic side view of MLG which rests on top of the $(6\sqrt{3} \times 6\sqrt{3})\text{R}30^\circ$ reconstructed ZLG which is weakly bonded to the underlying SiC. Due to the covalent bonding between the ZLG and the underlying SiC, the ZLG exhibits a characteristic buckling. \textit{Ab initio} calculations by Varchon \textit{et al.} \cite{varchon2008ripples} and Kim \textit{et al.} \cite{kim2008origin} predict vertical displacements between \SI{.34}{\angstrom} and \SI{1.2}{\angstrom}, while grazing incidence fast atom diffraction measurements by Zugarramurdi \textit{et al.} \cite{zugarramurdi2015determination} report an average corrugation of \SI{0.27}{\angstrom}. A value of \SI{0.8}{\angstrom} was used in Fig.~\ref{fig:Sample}(a) for visual clarity.

The sample was first checked with atomic force microscopy (AFM) [Fig.~\ref{fig:Sample}(b)] to determine the quality of the growth process. AFM shows the typical step-bunching observed in the growth process and reveals terraces of \SIrange[range-units=single]{2}{4}{\micro\metre} width. Further characterization was carried out after transferring the sample into an ultra-high vacuum (UHV) environment and degassing with a resistive heater at \SI{750}{\degreeCelsius} for 30 min. When observed with low-energy electron diffraction (LEED) [Fig.~\ref{fig:Sample}(c)], the $(6\sqrt{3} \times 6\sqrt{3})\text{R}30^\circ$ surface reconstruction of the ZLG can be seen as well as the first order diffraction spots of the MLG and SiC. LEED data were taken with the sample normal to the incident electron beam, using a beam energy of \SI{80}{\electronvolt}.

\subsection{Experimental procedure}

\begin{figure}[htbp]
    \centering
     \includegraphics[width=1\linewidth]{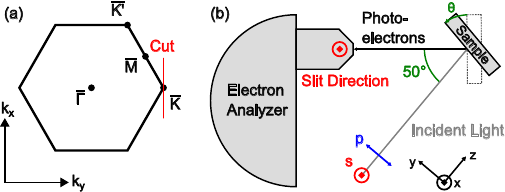}
    \caption{(a) Surface Brillouin zone of graphene in which the orientation of the $k$-space cuts for photon-energy dependent ARPES is indicated by the red line. (b) Sample orientation relative to the electron analyzer and the incident light. The analyzer slit direction is perpendicular to the plane of light incidence.}
    \label{fig:Geometry}
\end{figure}

The sample was transferred in air to beamline 5-2 (BL5-2) at the Stanford Synchrotron Radiation Light-source (SSRL) where it was loaded into UHV and degassed at \SI{300}{\degreeCelsius} for \SI{1}{\hour}. ARPES energy-momentum cuts were taken through the Dirac point at $\overline{\mathrm{K}}$, perpendicular to the $\overline{\Gamma\mathrm{K}}$ direction [Fig. \ref{fig:Geometry}(a)] using a Scienta Omicron DFS30 electron analyzer with a vertical entrance slit. The sample was held at \SI{150}{K} and a pressure was less than \SI{3e-11}{torr} for the duration of the measurements. Measurements were made using $p$ polarized light with photon energies in the range from \SIrange{20}{200}{\electronvolt} in \SI{2}{\electronvolt} steps. As sketched in Fig.~\ref{fig:Geometry}(b), the sample was rotated from $\theta \approx  \SIrange{55}{11}{\degree}$ relative to the spectrometer axis to remain at the $\overline{\mathrm{K}}$ point as the photon energy was changed. While the low photon-energy light was almost normal to the sample, the angle of incidence was gradually increasing as the sample was rotated towards the analyzer with increasing $h\nu$. For constant energy plots, an electron deflector was used as opposed to rotating the sample away from $\overline{\mathrm{K}}$.

\section{Computational Methods}

\subsection{Final-state plane wave and trLEED}

To simulate ARPES photoemission intensity, the electron dipole transition matrix elements were first computed in the nearly free electron or final-state plane wave (FSPW) approximation. As shown in Ref. \cite{moser_experimentalists_2017}, this model uses a single plane wave to model the final state of a photoelectron and is effective at capturing orbital characteristics and the geometric phase interference of electrons emitted from different atomic positions within the crystal lattice. The initial states were determined from a tight-binding model of unstrained MLG with first nearest neighbor hopping parameters reported in Ref.  \cite{rezaei2018modified}, alongside a monolayer work function of \SI{4.16}{eV} \cite{mammadov_work_2017}. Computations were performed using the \textit{chinook} python package by Day \textit{et al.} \cite{day_computational_2019}. The photoemission intensity was calculated using $p$-polarized light with a photon-energy step size of  \SI{1}{\electronvolt}.

In Day's implementation of the FSPW model, the dipole operator is implemented in the length gauge, and the matrix element for each orbital index $a$, $M_{a}^{\text{FSPW}}(\vec{k},\omega)$, is evaluated as a function of the electron wave vector $\vec{k}$ and incident photon energy $\omega$ using a spherical-harmonic expansion of the atomic orbitals \cite{day_computational_2019}. Because the final state is approximated as a plane wave  $e^{i\vec{k}\cdot\vec{r}}$, the dipole matrix element reduces to a Fourier transform of the initial state atomic orbitals. Since the Fourier transform of the spherical harmonics is known, the software uses a table of precalculated values to avoid the computationally expensive real-space integrals  \cite{day_computational_2019}. The total matrix element is expressed as a weighted sum of the precalculated terms which allows for the program to be very computationally efficient.

To approximate a more realistic final state while preserving computational efficiency, we included a simple model of time-reversed LEED (trLEED) states \cite{feibelman1974photo} that captures reflection of the photoelectron from the surface potential of the MLG. The trLEED final state inside a material has two components: a transmitted wave ($k_z>0$) and a reflected wave ($k_z<0$). To bypass full real-space $z$-integration, we approximate this state using a complex coefficient $\beta(E)$ that accounts for the resulting constructive and destructive interference. By matching the internal wave function with the external vacuum solution at the surface boundary ($z=0$) for an electron propagating through an effective monolayer depth $z_0$, we define
\begin{equation}
\beta(E) = e^{i k_{z} z_{0}} + \alpha\, e^{-i k_{z} z_{0}},
\end{equation}
\noindent where
\begin{equation}
\alpha(E) = \frac{k_{z} - k_{z}'}{k_{z} + k_{z}'}.
\label{eq:alpha}
\end{equation}
Here, the reflection amplitude $\alpha$, Eq. \eqref{eq:alpha}, is modeled as reflection from a step potential with complex inner potential $V = -V_0+iV_i$ and electron kinetic energy $E$ \cite{pendry1990low, pendry1976theory}, where $k_{z} = \frac{1}{\hbar}\sqrt{2 m_{e}(E \cos^2\theta + V_0)}$ represents the wave in the material and $k_{z}' = \frac{1}{\hbar}\sqrt{2 m_{e} E \cos^2\theta}$ represents the free wave in vacuum which behaves like $e^{ik_z'z}$. For our calculations $V_0$ and $V_i$ were treated as modeling parameters with initial values set to \SI{16.4}{\electronvolt} \cite{gruneis2008electron} and \SI{0}{\electronvolt} respectively. At $z_0 = 0$, this simplifies to the original FSPW model up to the energy-dependent reflection factor $1 + \alpha(E)$. 

This complex amplitude $\beta(E)$ acts as an overall phase and attenuation factor for the final state wave function. Since this trLEED modification is implemented by multiplying the plane-wave final state by $\beta(E)$, it simply scales each term in \textit{chinook's} precalculated sum. Expressing the kinetic energy as $E(\omega) = \hbar \omega - \phi - E_B$, the modified matrix element is written as
\begin{equation}
M_{a}^{\text{trLEED}}(\vec{k},\omega) = \beta(E(\omega)) M_{a}^{\text{FSPW}}(\vec{k},\omega),
\end{equation}
\noindent where $\phi$ is the material's work function and $E_B$ is the binding energy.

This approach successfully captures out-of-plane interference and phase-shifting effects without introducing real-space numerical integrals over $z$, completely retaining the original computational speed of the FSPW approximation. The resulting photoemission intensity is evaluated as
\begin{equation}
I(\vec{k},\omega) \propto \left| \sum_{a} M_{a}^{\text{trLEED}}(\vec{k},\omega) \right|^2.
\end{equation}

\subsection{Real-space multiple scattering}

\begin{figure*}[ht]
    \centering
    \includegraphics[width=1\linewidth]{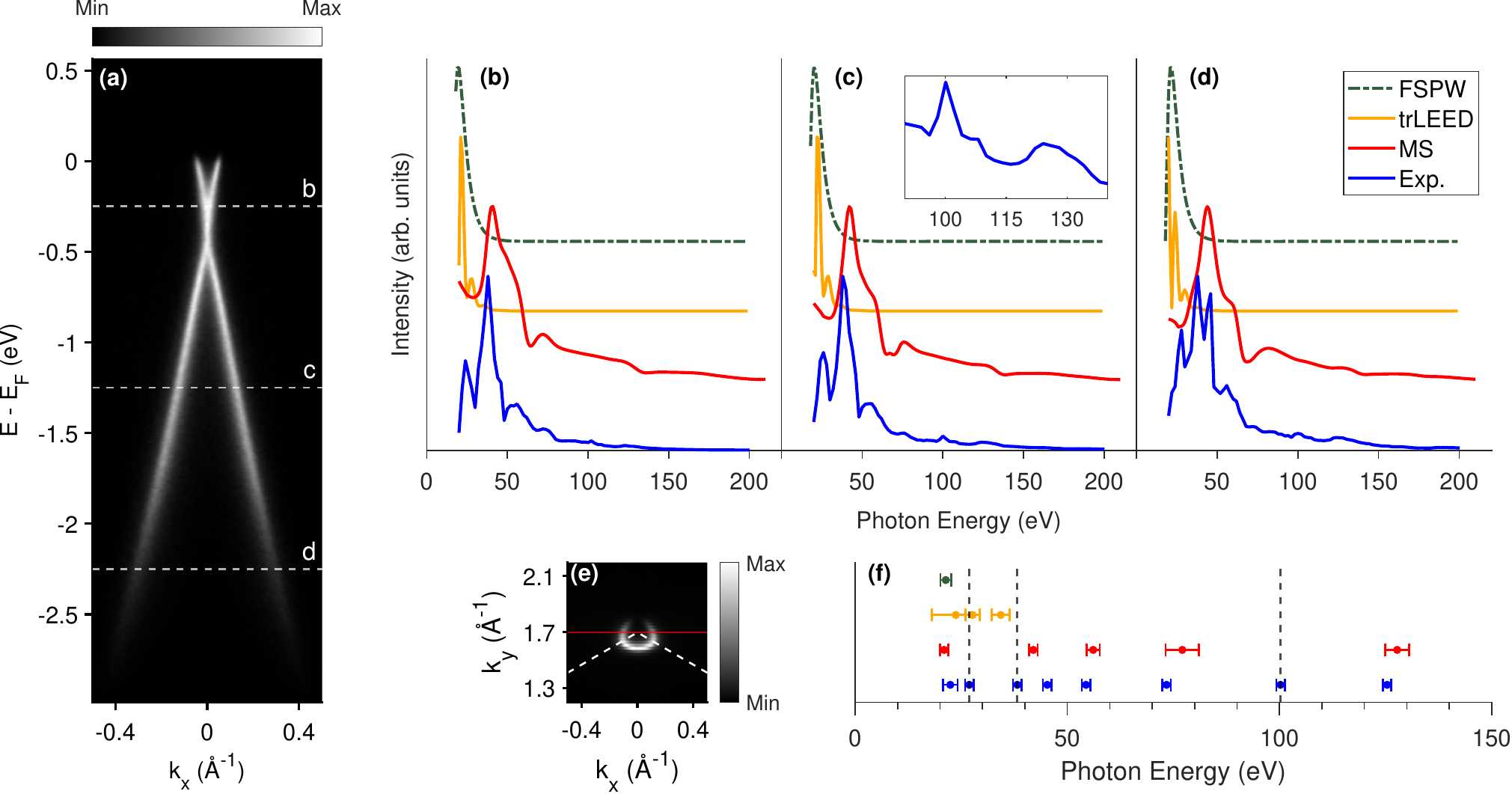}
    \caption{(a) Characteristic ARPES scan at  $h\nu = \SI{40}{\electronvolt}$ through the $\overline{\mathrm{K}}$ point with dashed lines at  $E - E_\text{F} =  \SIlist[list-units=single]{-0.25; -1.25; -2.25}{\electronvolt}$, indicating the binding energies used for the intensity plots in (b)-(d). (b)-(d) Photoemission intensity as a function of photon energy for experimental data, FSPW, trLEED, and MS simulations. For the experimental data, background intensity was removed as outlined in Appendix~\ref{mask}. Multiple Gaussian distributions were fitted to determine resonant peak positions for all but the FSPW model where a Lorentzian distribution was used. (e) ARPES constant energy surface taken at $E- E_\text{F} = \SI{-1.25}{\electronvolt}$, with a red line indicating the cut direction in (a) and dashed white lines indicating the boundaries of the first Brillouin zone. (f) Average photon energy position of the resonant peaks, determined from ten binding energy cuts. The error bars represent the standard deviation of the peak positions, with a minimum floor set to \SI{1}{\electronvolt}.}
    \label{fig:hvDependence}
\end{figure*}

In a second approach, the ARPES spectra were computed in the one-step model of photoemission with trLEED boundary conditions, using the real-space multiple scattering (MS) method by Krüger \textit{et al.} \cite{kruger_real-space_2011}. Unlike the FSPW and the simple trLEED approximations used in the previous section, this model describes the photoelectron wave as the sum of a plane wave and spherical waves incoming at each atomic site and relies on a finite-size cluster of atoms in contrast to the periodic in-plane boundary conditions used in \textit{chinook}. 

A free-standing MLG sheet was modeled with a disk-shaped cluster of 348 carbon atoms and 1185 empty cells. These empty cells are used for describing the spatial extension of C $s$ and $p$ orbitals beyond the radius of the C atomic spheres and are needed to properly represent the anisotropic graphene potential within the muffin-tin approximation \cite{kruger2022observation}. The potentials of the atomic carbon and empty spheres were determined for the infinite graphene layer using the linear-muffin-tin orbital software in the atomic sphere and local density approximation \cite{andersen1975linear}.

The matrix elements of the transition dipole operator are evaluated in acceleration gauge \cite{pendry1976theory}. In this form the dipole operator is approximately $\vec{\epsilon} \cdot \nabla V$, where $\vec{\epsilon}$ is the polarization vector of the incoming photon and $V$ is the crystal potential. This is a convenient choice in the muffin-tin approximation as inside the atomic spheres $\nabla V = \left( dV/dr \right) \hat{r}$ and $\nabla V = 0$ in the interstitial region \cite{kruger2018photoelectron}. Because of this, the interstitial region does not contribute to the transition matrix. This computational approach was shown to be very reliable for ARPES intensity calculations of freestanding MLG \cite{kruger2022observation, nozaki_computational_2024}. The photoemission intensity was calculated as a spherical tensor of the polarization vector, which were then used to reconstruct the experimental p-polarization. The calculations were performed with a photon-energy step size of \SI{1}{\electronvolt}.

\section{Results and Discussion}

\subsection{Out-of-plane effects on intensity resonances}  \label{3DScattering}

To probe substrate interactions, we utilized photon-energy scanned ARPES to resolve final-state scattering enhancements across the MLG/ZLG/SiC heterostructure. The orientation of our measurements relative to the MLG Brillouin zone is shown with dashed lines in the constant energy map of Fig.~\ref{fig:hvDependence}(e) and in Fig.~\ref{fig:Geometry}(a) with a solid red line through the $\overline{\mathrm{K}}$ point. Fig.~\ref{fig:hvDependence}(a) displays the corresponding $E$ vs. $k_x$ dispersion at $h\nu = \SI{40}{\electronvolt}$, where the Dirac bands of the MLG are resolved. The horizontal dashed lines in Fig.~\ref{fig:hvDependence}(a) indicate three representative binding energies at which the intensity was tracked as a function of $h\nu$ as presented in Fig.~\ref{fig:hvDependence}(b)-(d). To isolate the spectral weight of the $\pi$ bands, a numerical mask was used (Appendix~\ref{mask}). This captures the total cross section of the dispersing bands, excluding the background without distorting the lineshape or intensity. The intensity values shown in Fig.~\ref{fig:hvDependence}(b)-(d) represent the $k_x$-integrated sum of the masked signal. This extraction was completed systematically across ten binding energies, three of which are highlighted here.

The key result of this paper is the multiple matrix element modulations observed in the plot of photoemission intensity with respect to $h\nu$, as shown by the blue curves in Fig.~\ref{fig:hvDependence}(b)-(d). These photoelectron intensity modulations are different from the so-called \emph{scattering resonances} originating from initial state effects that do not depend on photon energy \cite{nazarov2013scattering, krivenkov2017suppression, matta2022momentum}. The precision of the extracted resonant energy positions from the experimental data in Fig.~\ref{fig:hvDependence}(b)-(d) is limited by the \SI{2}{\electronvolt} photon-energy step size. This may result in slight shifts in the extracted peak positions; however, we do not expect these to significantly affect the conclusions drawn from their locations.

\begin{figure}[t]
    \centering
    \includegraphics[width=1\linewidth]{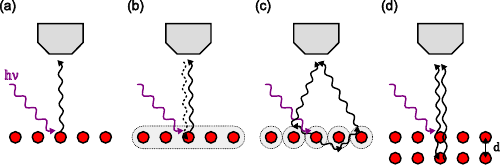}
    \caption{Cartoons of the (a) final-state plane wave approximation, (b) time-reversed LEED state, (c) real-space multiple scattering, and (d)  interlayer interference. The red circles represent the C atoms, and the black wavy arrow represents the emitted photoelectron. In (a), we see an electron that does not interact before reaching the detector. (b) shows an electron and its time-reversed LEED state which can interfere before reaching the detector. The gray region represents the inelastic mean free path of the electron within the MLG. In (c), the electron can scatter (elastically) off neighboring muffin-tin potentials (gray circle) as implemented in the real-space MS model \cite{kruger_real-space_2011}. (d) shows two electrons emitted from the same location scatter off two different layers of the system. This results in interlayer interference that is sensitive to the kinetic energy of the electrons.}
    \label{fig:modelDiagram}
\end{figure}

To determine the physical mechanism that is responsible for these enhancements, we first implemented a FSPW simulation of photoemission from a freestanding monolayer of graphene [black dashed line in Fig.~\ref{fig:hvDependence}(b)-(d)], as schematized in Fig.~\ref{fig:modelDiagram}(a). As can be seen in Fig.~\ref{fig:hvDependence}(f) this model replicates the first peak. This method also correctly captures orbital characteristics such as the dark corridor \cite{nozaki_computational_2024}, but fails to capture any other $h\nu$-dependent features. The failures of this model are consistent with past results \cite{nozaki_computational_2024, dauth_perpendicular_2016, ono2018one, kern_simple_2023}.

Next, we tested a simple trLEED extension to the model, Fig.~\ref{fig:modelDiagram}(b), which includes a wave component traveling into the monolayer. By modifying the FSPW model to include this reflected wave component, the interference between the incoming and outgoing waves is captured with a single phase factor. The results of these calculations are shown in orange in Fig.~\ref{fig:hvDependence}(b)-(d). The trLEED model is able to qualitatively reproduce the first three peaks when using an effective emission depth of $z_0 = \SI{6}{\angstrom}$ and an inner potential of $V_0 = \SI{16.5}{\electronvolt}$ \cite{razado2018structural}. Within this model the locations of the second and third peaks can be shifted to higher or lower $h\nu$ by varying the inner potential. The number of visible peaks can also be increased by increasing the effective emission depth. However, the additional peaks are periodic which does not reflect the experimental results. Furthermore, the model requires a large effective emission depth of \SI{6}{\angstrom} to even qualitatively match the experimental results, which is physically inconsistent with the thickness of a freestanding graphene monolayer (\SI{3.35}{\angstrom}) \cite{shearer2016accurate}. Ultimately this model fails to reproduce the experimental results but shows that the addition of a reflected wave and finite $z_0$ component of electron propagation can introduce intensity oscillations.

Since the simple models fail to capture many of the features observed in the experiment, a more sophisticated approach is necessary. A MS model  \cite{ kruger_real-space_2011} is implemented for freestanding monolayer graphene. In this model, Fig.~\ref{fig:modelDiagram}(c), the emitted electrons are scattered off neighboring atomic potentials. The resulting phase shifts allow for new constructive interference conditions that produce additional $h\nu$-dependent intensity peaks. The MS calculation, as presented in the red curves of Fig.~\ref{fig:hvDependence}(b)-(d) accurately predicts the location of many of the peaks found in the experiment data [Fig.~\ref{fig:hvDependence}(f)], indicating that in-plane scattering of photoelectrons within the graphene lattice [Fig.~\ref{fig:modelDiagram}(c)] is the primary perturbation of the matrix element in this energy range. The polarization dependence of the calculated intensity resonances is also investigated using the MS simulations, with comparisons for $s$-polarized and circularly polarized light provided in Appendix~\ref{polarization}. Fig.~\ref{fig:hvDependence}(f) shows that the MS model effectively tracks the experimental peak positions with a few notable disagreements: there are three experimentally observed peaks at $h\nu = \SIlist[list-units=single, separate-uncertainty=true]{27 \pm 1; 38 \pm 1; 100 \pm 1}{\electronvolt}$ that are not present in the MS data. The difference between the experimental data, the MS model data and the large emission distance ($z_0 = \SI{6}{\angstrom}$) needed for our trLEED model strongly indicates that the source of these additional final-state scattering enhancements must originate from out-of-plane interactions that are not captured by the freestanding graphene models. 

With freestanding MLG simulations unable to reproduce all observed resonances, we turn to a structural analysis of the heterostructure. To determine if these additional peaks come from interlayer photoelectron interference [Fig.~\ref{fig:modelDiagram}(d)], we map the experimental data from $h\nu$ to the electron wave vector $k$. This is done using the standard free-electron final state approximation accounting for the inner potential $V_0$, binding energy $E_{B}$, and work function $\phi = \SI{4.16}{\electronvolt}$ \cite{mammadov_work_2017}. Here the wavevectors $k_i$ inside the material and $k_o$ outside the material are given by
\begin{equation*}
 k_{i} = \frac{\sqrt{2m_{e}(h\nu-\phi-E_B + V_0)}}{\hbar}
\end{equation*}
\noindent  and,
\begin{equation*}
 k_{o} = \frac{\sqrt{2m_{e}(h\nu-\phi-E_B)}}{\hbar},
\end{equation*}
\noindent  with the corresponding resonance condition expressed as
\begin{equation*}
2k_{i}x_1 – k_{o}x_2 = 2\pi n,
\end{equation*}
\noindent where $x_{1} = {d}/ \cos \theta_{i}$ and $x_{2} = 2{d} \sin \theta_{o} \tan \theta_{i}$. Here, $d$ is the vertical separation between two lattice planes and $\theta_{i}$ and $\theta_{o}$ represent the angle of the electron propagating relative to the surface normal inside and outside of the material. This can then be further simplified to,
\begin{equation}
2d k_i \cos\theta_i = 2\pi n. %\label{eq:interlayer_phase}
\end{equation}
\noindent The derivation of this expression is detailed in Appendix \ref{derivation}.

For our calculations, $V_0$ was optimized to \SI{15.95}{\electronvolt}, which is close to the expected value for MLG \cite{razado2018structural}. Now applying the above condition, we find that the three experimental peaks at $h\nu = \SIlist[list-units=single, separate-uncertainty=true]{27 \pm 1; 38 \pm 1; 100 \pm 1}{\electronvolt}$ correspond to vertical distances of $d = \SIlist[list-units=single, separate-uncertainty=true]{3.49 \pm 0.06; 1.97 \pm 0.03; 0.61 \pm 0.003}{\angstrom}$, respectively. These values correspond to interference conditions for the MLG/ZLG separation, the ZLG/SiC interfacial distance, and the short internal C-Si spacing within the SiC bilayers. Our results are in excellent agreement with previous structural determinations by Razado-Colambo \textit{et al.} \cite{razado2018structural}. Using synchrotron-radiation photoelectron diffraction they found that the corresponding distances were $d = \SIlist[list-units=single]{3.48;2.0; 0.63}{\angstrom}$. This provides strong evidence that these features originate from the 3D geometry of the multilayer stack rather than the 2D structure of graphene. 

If, instead, $V_0$ is extracted using the known distances reported by Razado-Colambo \textit{et al.} \cite{razado2018structural}, the inner potential for each successive scattering event would be approximately $V_0 \approx \SIlist[list-units=single]{16.2; 14.8; 10.0}{\electronvolt}$. This trend indicates an effective inner potential that decreases monotonically when moving spatially deeper from the MLG surface, through the ZLG interface, and into the SiC substrate. While $\SI{16.2}{\electronvolt}$ and $\SI{14.8}{\electronvolt}$ are well within the expected range for MLG \cite{razado2018structural} and the MLG/ZLG/SiC interface, the lowest value ($\SI{10.0}{\electronvolt}$) is smaller than typically expected for a bulk semiconductor \cite{kruse2003determination, kruse2006determination, pennington2010theoretical}.

The present interference model could be extended to other 2D materials and heterostructures, including intercalated quasi-free-standing graphene (QFMLG), hexagonal boron nitride, and transition-metal dichalcogenides (TMDs). Beyond the determination of interlayer separations, the significance of this approach lies in understanding how out-of-plane structural interactions influence the photoemission response of electronic states. For example, applying this framework to intercalated QFMLG or other vdW heterostructures could help distinguish changes in the intrinsic electronic structure associated from intensity modulations arising from final-state scattering. Extending the model to incorporate element-dependent scattering phase shifts and realistic interface electronic states would be an important direction for future studies, particularly for materials containing heavier elements such as TMDs.

Regarding the resonance observed at $h\nu = \SI{100}{{\electronvolt}}$, an experimental artifact from the beamline diffraction grating causes the Si $2p$ core level to appear in the background of the ARPES scan. While this partially enhances the intensity observed at $h\nu = \SI{100}{\electronvolt}$, we note that the resonance persists at binding and photon energies isolated from this contribution. A detailed analysis of this artifact is presented in Appendix~\ref{XPS}.

\subsection{Photon-energy dependence of replica bands} \label{Replica}

\begin{figure}[tbp]
    \centering
    \includegraphics[width=1\linewidth]{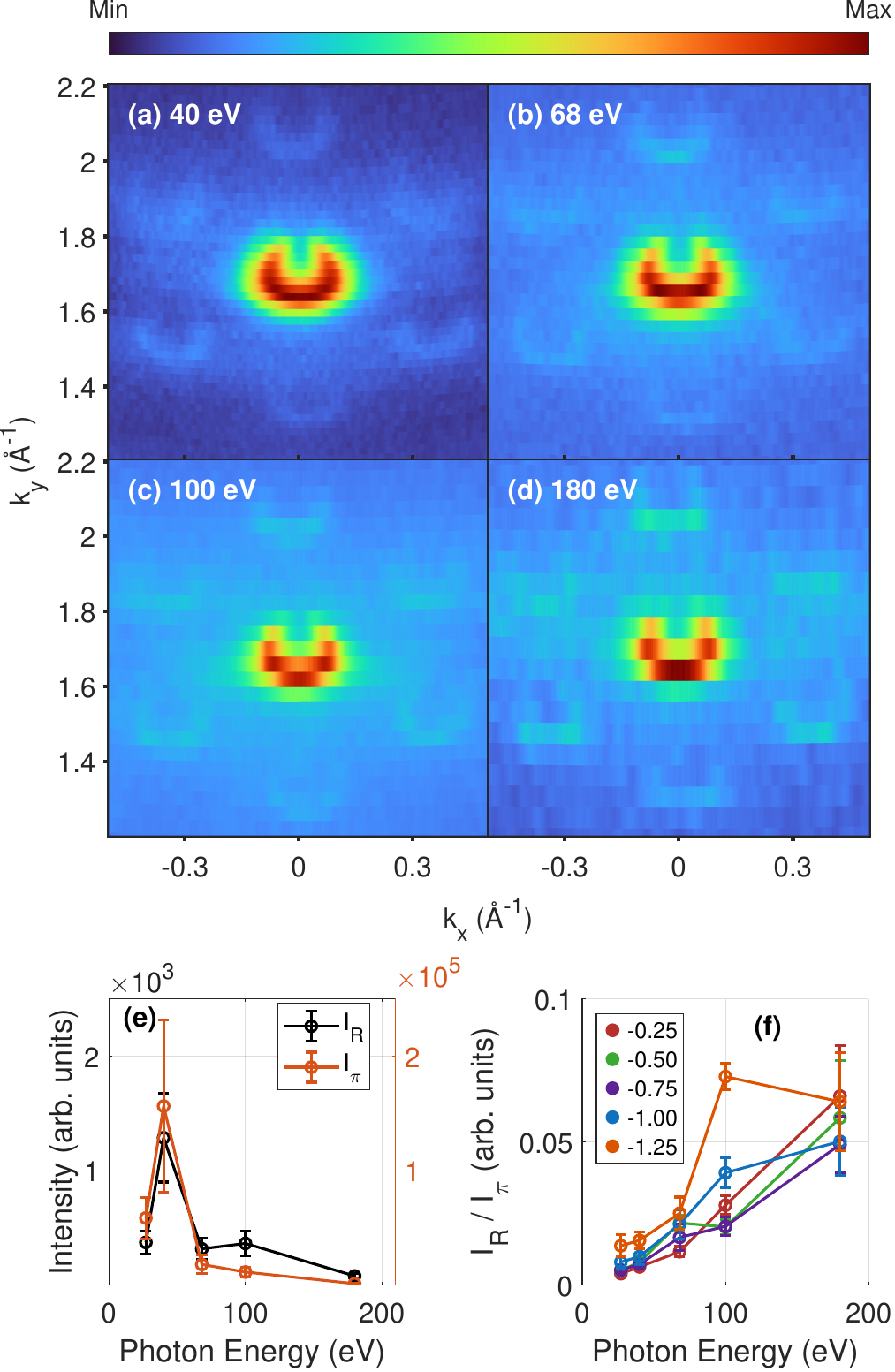}
    \caption{(a)-(d) Constant energy maps taken at $E - E_\text{F} = \SI{-1.00}{\electronvolt}$, for $h\nu = \SIlist[list-units=single]{40; 68; 100; 180}{\electronvolt}$. (e) Comparison of $I_{\pi}$ and $I_R$ using the average of the intensities for five different energies $E - E_\text{F}$ at each photon energy. (f) Ratio of the replica-band intensity ($I_R$) to the $\pi$-band intensity ($I_{\pi}$) as a function of photon energy for cuts through the $\overline{\mathrm{K}}$ point perpendicular to the $\overline{\Gamma\mathrm{K}}$ direction at various $E - E_\text{F}$. Intensities at each photon energy were normalized using the synchrotron beam flux and momentum-space maps were acquired using an electrostatic electron deflector.}
    \label{fig:replicaBand}
\end{figure}

\begin{figure*}[htbp]
    \centering
    \includegraphics[width=1\linewidth]{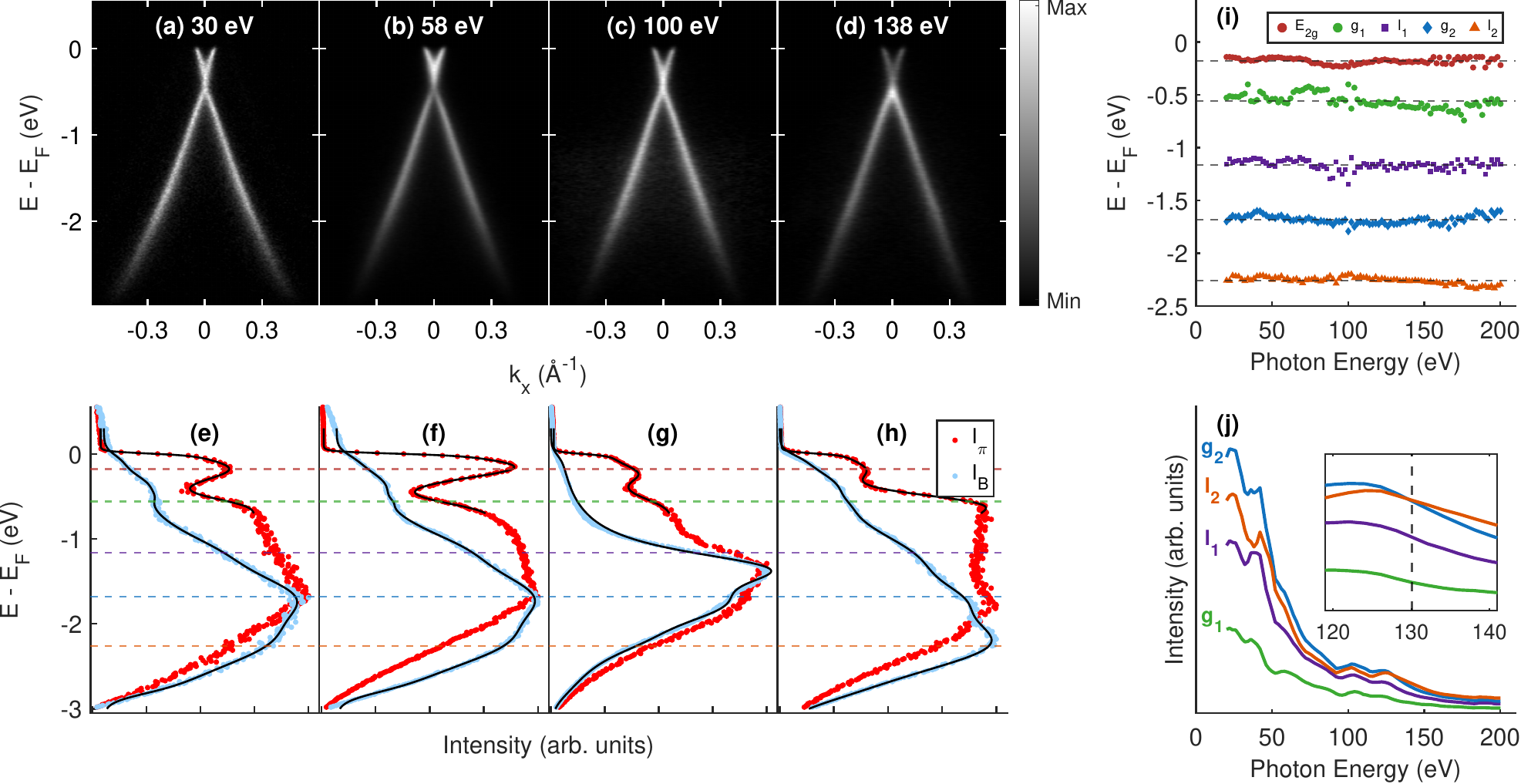}
    \caption{(a)-(d) ARPES scans of the Dirac cone taken at $h\nu = \SIlist[list-units=single]{30; 58; 100; 138}{\electronvolt}$. (e)-(h) Integrated intensity as a function of $E - E_\text{F}$ for the $\pi$ bands ($I_{\pi}$, red) and the background signal ($I_B$, blue) corresponding to panels (a)-(d). Solid black curves are fits to the data. (i) Tracking of the energy levels ($E - E_{\text{F}}$) for the peak near the $E_{2g}$ EPC kink and interfacial states $g_1$, $l_1$, $g_2$, and $l_2$ across the measured photon energy range. Dashed lines represent the mean binding energy for each state. (j) Photoemission intensity as a function of photon energy for the four interfacial states.}
    \label{fig:EBDependence}
\end{figure*}

Fig.~\ref{fig:replicaBand}(a)-(d) shows constant energy maps around the $\overline{\mathrm{K}}$ point of the MLG Brillouin zone \SI{1.00}{\electronvolt} below the Fermi level. Here the characteristic \emph{dark corridor}, where matrix-element effects suppress photoemission in certain directions, can be seen. Surrounding the primary Dirac cone contour are the $(6\sqrt{3} \times 6\sqrt{3})\text{R}30^\circ$ replica bands. The physical origin of these replicas has been attributed to two competing models. These are photoelectron final-state scattering \cite{polley2019origin} and initial-state modulation because of the real-space moiré periodicity coming from the ZLG and SiC substrate \cite{huang2017effects}. Our results show a scaling of the replica-band intensity that is dependent on $h\nu$. To our knowledge, this scaling behavior has not been previously reported and strongly favors the final-state scattering interpretation.

If the replica-bands were purely an initial-state effect, then we would expect that their intensity is coupled to that of the $\pi$ band, maintaining a constant intensity ratio regardless of $h\nu$. Fig.~\ref{fig:replicaBand}(e) shows that the absolute magnitude of both the replica-band intensity ($I_R$) and the $\pi$ band intensity ($I_{\pi}$), averaged across five binding energies, do follow the same overall trend as the intensities in Fig.~\ref{fig:hvDependence}(b)-(d). However, as seen in Fig.~\ref{fig:replicaBand}(f), the ratio of the replica band intensity to the $\pi$ band intensity ($I_{R}/I_{\pi}$) increases over the $h\nu$ energy range from \SIrange{27}{180}{\electronvolt}. This shows that the replicas experience a less steep decrease in relative intensity than the primary $\pi$ band. We note the outlying data at $E - E_\text{F} = \SI{-1.25}{\electronvolt}$, $h\nu = \SI{100}{\electronvolt}$ in Fig.~\ref{fig:replicaBand}(f). This comes from photoemission of the Si $2p$ core level that was discussed previously, cf. Appendix~\ref{XPS}. Here the increased background intensity artificially increases the replica band signal. 

To determine the source of the energy dependence of the intensity ratio $I_{R}/I_{\pi}$, we first consider the contribution of the IMFP and the polarization dependence of the matrix element. Historically, the IMFP was treated as a universal curve, described by Seah and Dench \cite{ seah1979quantitative}, with a minimum near $E = \SI{50}{\electronvolt}$. For MLG, modern determinations of the IMFP range from approximately constant for energies up to \SI{100}{\electronvolt}, to monotonically decreasing for energies up to \SI{25}{\electronvolt} \cite{nguyen2020low, geelen2019nonuniversal}. Neither trend is consistent with the monotonic increase in the ratio observed here, indicating that the origin is not rooted in simple attenuation of the electron propagating through the heterostructure. Secondly, we can distinguish this behavior from the polarization-dependence of the matrix element. In our experimental geometry, the sample is rotated as $h\nu$ is varied, which results in a change in the angle between the light's polarization vector and the sample surface normal. While small differences in momentum-space position and orbital character may result in changes to the polarization-dependence of the photoemission matrix elements, making it impossible to entirely exclude these effects as a contribution to the observed intensity ratio, we expect their impact to be relatively minor given the similar momentum-space locations and orbital character of the primary and replica bands. Instead, the observed behavior may depend on how the scattering cross section changes in the experiment at different $h\nu$.

For electrons scattered by atomic potentials, the scattering cross section is dependent on both the kinetic energy of the electron and the scattering angle $\theta_s$, where $\theta_s = \SI{180}{\degree} - 2 \theta_i$. As shown by Heinz, the magnitude of the atomic scattering factor $\vert t(k, \theta_s) \vert$ (where $k \propto \sqrt{E}$) increases with $E$ \cite{heinz2013leed}. Further, $\vert t(k, \theta_s) \vert$ for C atoms exhibits maximum values in the forward $(\theta_s = \SI{0}{\degree})$ and backward $(\theta_s = \SI{180}{\degree})$ scattering directions and a minimum near $\theta_s \approx \SI{90}{\degree}$ \cite{heinz2013leed}. Crucially, $\vert t(k, \theta_s) \vert$ for C increases monotonically with $\theta_s$ while moving from the minimum toward the backscattering direction \cite{heinz2013leed}.

In our experiment, the sample angle is adjusted with $h\nu$ to align photoemission from the $\overline{\mathrm{K}}$ point with the analyzer axis. At lower photon energies ($\SI{27}{\electronvolt}$), emission occurs at higher off-normal angles, corresponding to a ZLG scattering angle of $\theta_s \approx \SI{108}{\degree}$ near the minimum of the final-state scattering cross section. As $h\nu$ increases, the emission direction rotates closer to normal incidence, driving $\theta_s$ toward $\SI{180}{\degree}$ (perfect backscattering). Consequently, the overall scattering rate increases at higher $h\nu$ due to both the increase in $E$ and the angular trajectory shift toward stronger backscattering. While these results cannot completely exclude contributions from the initial state, they do indicate that $I_{R}/I_{\pi}$ (and hence the presence of the replica bands) is predominantly driven by final-state backscattering of the photoelectrons.

\subsection{ZLG interfacial states and $\pi$-band electron phonon coupling} \label{zlgSurface}

To disentangle the diverse electronic signatures within the MLG photoemission spectra, we now examine the summed energy distribution curves for the $\pi$ bands and the background signal across all $81$ $h\nu$ scans. By separating these components, we identify localized electronic states at the ZLG/SiC interface within the background signal, alongside a distinct feature in the $\pi$ band signal that coincides with the known $E_{2g}$ electron-phonon coupling (EPC) signature \cite{bostwick2007quasiparticle, calandra2007electron}. Figure~\ref{fig:EBDependence}(a)-(d) shows the MLG $\pi$ bands at $h\nu = \SIlist[list-units=single]{30; 58; 100; 138}{\electronvolt}$. To better distinguish the dispersing bands from the background, we utilized the masking procedure outlined in Appendix~\ref{mask} to extract the momentum-integrated intensity for the $\pi$ bands ($I_{\pi}$) and the background ($I_B$) [Fig.~\ref{fig:EBDependence}(e)-(h)].

Turning first to the extracted $\pi$-band intensity ($I_\pi$) shown in Fig.~\ref{fig:EBDependence}(e)-(h), we observe a notable maximum near the energy of the known $E_{2g}$ EPC kink \cite{bostwick2007quasiparticle, calandra2007electron}. The peak location was determined by fitting the data with a model consisting of a primary Lorentzian peak and a Gaussian component (accounting for the intensity increase at higher binding energies), multiplied by a Fermi-Dirac distribution, plus a constant background. Figure~\ref{fig:EBDependence}(i) shows that the location of the peak remains constant across all measured photon energies. The average value of $E - E_\text{F} \approx \SI{-.2}{\electronvolt}$ is in excellent agreement with previous theoretical and experimental results \cite{bostwick2007quasiparticle, calandra2007electron}. We note that the relative intensity of the peak is sensitive to the momentum-space alignment of the ARPES scan. The high intensity shown in Fig.~\ref{fig:EBDependence}(e) and (f), and the low intensity in (h) can be attributed to a small misalignment of the emission angle $\theta$ relative to $\overline{\mathrm{K}}$ for these photon energies.

Shifting our focus to the isolated background signal ($I_B$), our scans resolve four distinct electronic states localized at the interface, labeled $g_1$, $l_1$, $g_2$, and $l_2$ in Fig.~\ref{fig:EBDependence}(i). These interface features were fitted at each photon energy using four Lorentzian distributions alongside two background shaping components. A broad Gaussian centered near $E - E_\text{F} \approx \SI{-0.3}{\electronvolt}$ was used to capture intensity beyond $E_\text{F}$, which likely originates from higher-harmonic leakage through the monochromator, and a sigmoidal step function near $E - E_\text{F} \approx \SI{-2.9}{\electronvolt}$ to model the sharp drop in intensity at the edge of the detector energy window. A constant offset term was also included in the fit. As indicated by the dashed lines in Fig.~\ref{fig:EBDependence}(i), these features remain essentially fixed in binding energy across the entire measured photon energy range. The lack of $k_z$ dispersion confirms their origin as intrinsic initial states of the $(6\sqrt{3} \times 6\sqrt{3})\text{R}30^\circ$ reconstruction.

The states $g_1$ and $g_2$, located at $E - E_\text{F} \approx \SI{ -0.6}{\electronvolt}$ and $\approx \SI{-1.7}{\electronvolt}$, are consistent with previously reported surface states associated with partially covalent C-Si bonding \cite{emtsev2008interaction, qi2010epitaxial, nevius2015semiconducting, hwang2016study, n2017band}. The two additional states, which we have named $l_1$ and $l_2$, are located at $E - E_\text{F} \approx \SI{-1.2}{\electronvolt}$ and $\approx \SI{-2.3}{\electronvolt}$, respectively. Sensitivity of these binding energies to fitting constraints and Si $2p$ core-level contamination near $h\nu = \SI{100}{\electronvolt}$ introduced no observable change within our experimental resolution. These additional nondispersing states likely originate from the same ZLG/SiC interface. Density functional theory (DFT) calculations of the ZLG using the meta-generalized gradient approximation report multiple flat bands in the energy range of $E – E_\text{F} =\SIrange{-1.5}{0}{\electronvolt}$ \cite{mohebbi2025reliable}, while separate DFT calculations show a peak in the density of states at $E – E_\text{F} \approx \SI{-2.2}{\electronvolt}$ \cite{cavallucci2018intrinsic}, which is in good agreement with our observed $l_2$ state. 

After establishing their binding energies, we tracked the photoemission intensity of $g_1$, $l_1$, $g_2$, and $l_2$ as a function of photon energy $h\nu$ [Fig.~\ref{fig:EBDependence}(j)] to deduce their orbital character. Although the intensity modulation features described in Sec.~\ref{3DScattering} are present, they do not produce strong resonances here, manifesting instead as minor modulations superimposed on a monotonic decay. The states $l_1$, $g_2$, and $l_2$ display a sharp intensity drop from $h\nu \approx \SIrange{20}{80}{\electronvolt}$, closely mirroring the atomic photoionization cross section of C $2p$ orbitals \cite{yeh1985atomic}. From this behavior, we conclude, in agreement with Emtsev \textit{et al.} \cite{emtsev2008interaction}, that $g_1$ and $g_2$ originate predominantly from C $2p_z$ orbitals, and that this orbital character extends to $l_1$ and $l_2$ as well. Above $h\nu \approx \SI{130}{\electronvolt}$, the intensity of $l_2$ subtly overtakes that of $g_2$ [inset of Fig.~\ref{fig:EBDependence}(j)]. This intensity crossover aligns with the photoionization cross section of Si $3s$ and $3p$ orbitals, which surpass C $2p$ in this photon energy region \cite{yeh1985atomic}, suggesting partial hybridization with the underlying Si substrate.

\section{Conclusions}

In summary, we have comprehensively studied the electronic and structural properties of epitaxial graphene on SiC by means of photon-energy-dependent ARPES. By utilizing the large photon energy range provided by a synchrotron light source, we have resolved multiple interactions between the MLG and the substrate. First, we observed multiple $h\nu$-dependent resonances in the photoemission spectrum of the graphene $\pi$ bands and successfully reproduced most of these features with a MS model of freestanding MLG. We then demonstrated that electron final-state scattering from the substrate interfaces produces additional photon-energy resonances at \SIlist{27; 38; 100}{\electronvolt}. Secondly, we provided evidence that the $(6\sqrt{3} \times 6\sqrt{3})\text{R}30^\circ$ replica bands show intensity scaling governed by the final-state scattering dependence on $h\nu$. Lastly, we report the observation of four interfacial electronic states ($g_1, l_1, g_2, l_2$). Our study demonstrates that the out-of-plane structure of a heterostructure can strongly influence the measured photoemission intensity of its electronic states through final-state scattering. Extending this framework to other 2D materials and heterostructures could provide a means of disentangling intrinsic electronic-structure modifications from substrate-induced photoemission effects, facilitating a more reliable interpretation of ARPES measurements in complex layered systems.

\begin{acknowledgments}
This work is supported by the MonArk NSF Quantum Foundry from the National Science Foundation under Award No. DMR-1906383, and $\mu$-ATOMS Energy Frontier Research Center by the U.S. Department of Energy under Award No. DE-SC0023412. The graphene preparation was supported by the Deutsche Forschungsgemeinschaft (DFG, German Research Foundation) within the Research Unit FOR5242 through Projects No. Ku4228/1-1 and No. Sta315/13-1. Use of the Stanford Synchrotron Radiation Lightsource, SLAC National Accelerator Laboratory, is supported by the U.S. Department of Energy, Office of Science, Office of Basic Energy Sciences under Contract No. DE-AC02-76SF00515. E.F. and S.W. are supported via a fellowship by the NSF under the Award No. 2244274.
\end{acknowledgments}

\section*{Data Availability}
All raw data supporting the findings of this work are available from the corresponding authors upon reasonable request.

\appendix

\appsection{Numerical Masking Procedure} \label{mask}

\begin{figure}[h!]
    \centering
    \includegraphics[width=1\linewidth]{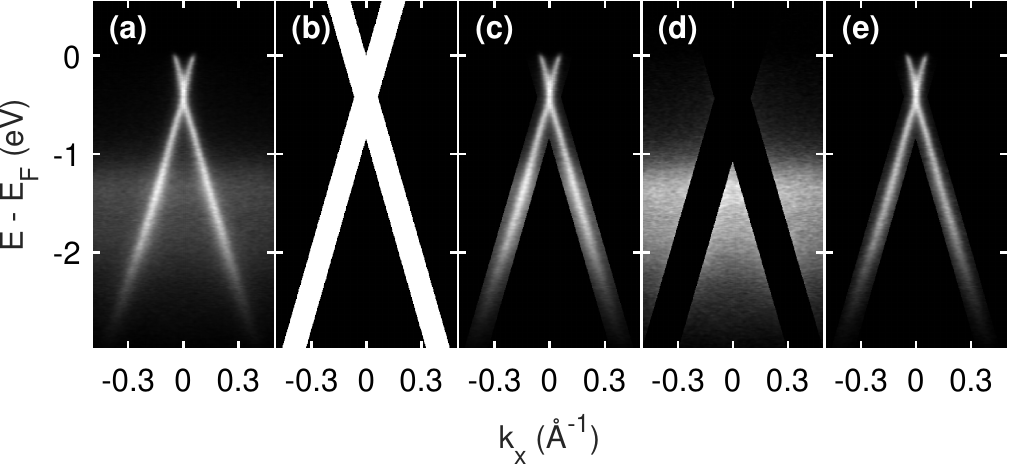}
    \caption{(a) Raw data taken at $h\nu = \SI{100}{\electronvolt}$. (b) Mask generated using the band dispersion. (c) Data with background removed using the mask. (d) Data with background only. (e) Data with $k_x$-averaged background subtracted.}
    \label{fig:Mask}
\end{figure}

Figure~\ref{fig:Mask} provides an example of the masking procedure used to isolate the electronic band dispersions from the nondispersing background signal. This approach was selected over derivative-based methods, such as the one by Zhang \textit{et al.} \cite{zhang2011precise}, which fail to maintain intensity information upon isolating dispersive features.

The mask was made in four steps. The first was to identify the two peak positions on either side of the Dirac point in the momentum distribution curve (MDC) for each energy $E - E_\text{F} < 0$. If no peaks greater than four times the signal-to-noise ratio were found, then that MDC was skipped. Next, the peak positions were used to fit two symmetric lines of the form $E-E_\text{F} = \pm v(k_x - k_0) + E_D$, where $(k_0, E_D)$ represents the coordinates of the Dirac point. Then, the mask was defined by setting all points within a total momentum width of $\SI{0.065}{\per\angstrom}$ (i.e., $\pm \SI{0.0325}{\per\angstrom}$ centered on the fitted lines) to 1 and 0 elsewhere [Fig.~\ref{fig:Mask}(b)]. The width of the region was chosen to avoid any clipping of the bands [Fig.~\ref{fig:Mask}(c)]. Finally, this mask was used to determine regions of background intensity [Fig.~\ref{fig:Mask}(d)] which were $k_x$-averaged for each energy and subtracted from the raw data [Fig.~\ref{fig:Mask}(e)]. A mask width of $\SI{0.065}{\per\angstrom}$ was applied to all data evaluated in Sec.~\ref{zlgSurface} for tracking the ZLG surface states, whereas a wider mask width of $\SI{0.1}{\per\angstrom}$ was used for all datasets in Sec.~\ref{3DScattering} to ensure no leakage of background intensity across the photon-energy-dependent analysis.

\appsection{Diffraction Grating Artifacts} \label{XPS}

When using synchrotron radiation for ARPES, a diffraction grating is used to monochromatize the photon source which allows for tunable photon energies. An unavoidable consequence of this is the leakage of higher-order diffraction peaks that will overlap the chosen $h\nu$. An example of this would be setting the monochromator to $h\nu = \SI{100}{{\electronvolt}}$ which will also pass the higher harmonics, such as $h\nu = \SI{200}{{\electronvolt}}$. The result of this leakage will typically be an energy shifted replica of the ARPES signal that appears at higher kinetic energies not detected during a measurement. In this study, however, the second harmonic of $h\nu = \SI{100}{{\electronvolt}}$ excites the Si $2p$ core level which appears in the background of the valence-band ARPES scan. An example of this can be found in Fig.~\ref{fig:EBDependence}(g) and Figs.~\ref{fig:Mask}(a) and \ref{fig:Mask}(d). Below we provide x-ray photoemission spectroscopy (XPS) data which confirms the location of the Si $2p$ core level.

\begin{figure}[htbp]
\centering
\includegraphics[width=1\linewidth]{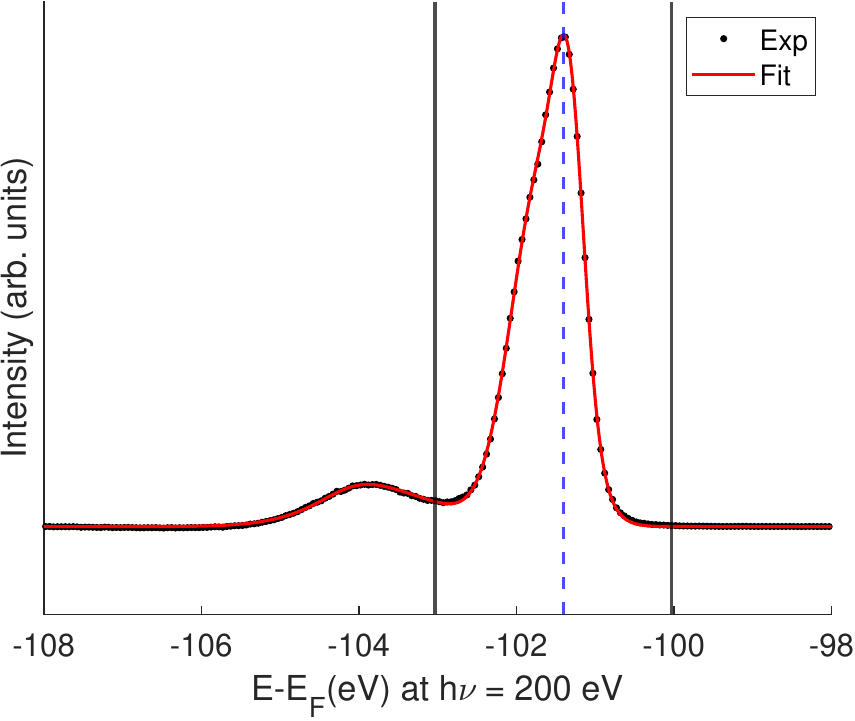}
\caption{XPS data of the Si $2p$ level taken at $h\nu = \SI{200}{{\electronvolt}}$ and normal emission. The solid black lines indicate the position of the intensity peak relative to the $h\nu = \SI{100}{{\electronvolt}}$ ARPES energy window where the right line is the Fermi edge and the left is $E-E_\text{F} = \SI{-3}{{\electronvolt}}$. The blue dashed line indicates the intensity maximum of the peak.}
\label{fig:XPS}
\end{figure}

Figure~\ref{fig:XPS} presents XPS data of the Si $2p$ core level taken at $h\nu = \SI{200}{\electronvolt}$. Here we observed two Si peaks coming from chemical bonding to oxygen ($E-E_\text{F} \approx \SI{103.7}{\electronvolt}$) and carbon ($E-E_\text{F} \approx \SI{101.3}{\electronvolt}$). The oxygen peak indicates partial undergrowth on some regions of the sample. These values are consistent with results from Refs.~\cite{chastain1992handbook, lu1993sio2}. When these core levels are excited by the second harmonic of $h\nu = \SI{100}{{\electronvolt}}$, the primary Si-C peak will appear at $E-E_\text{F} \approx \SI{-1.4}{{\electronvolt}}$ which corresponds to the background peak in Fig.~\ref{fig:EBDependence}(g).

\begin{figure}[htbp]
\centering
\includegraphics[width=1\linewidth]{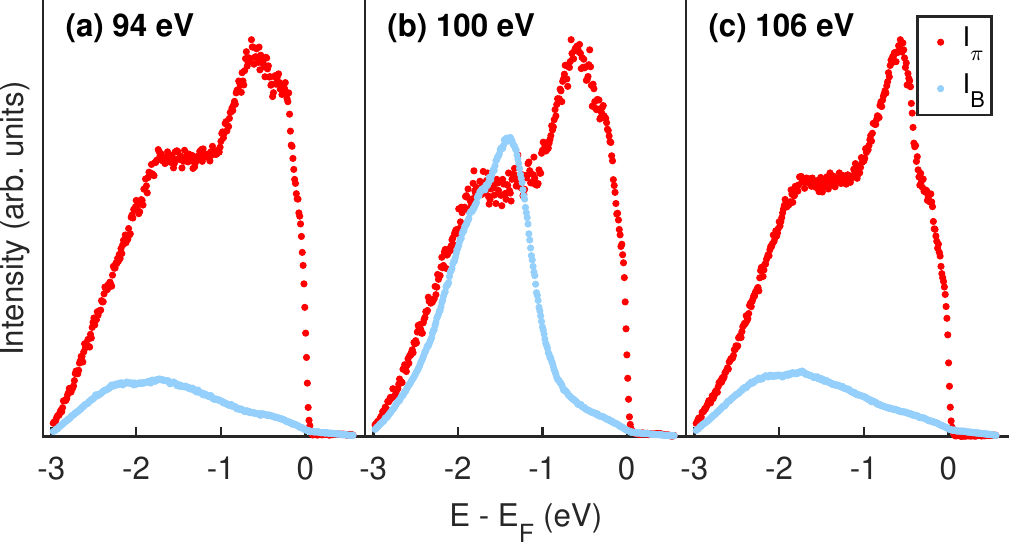}
\caption{(a)-(c) Intensity as a function of $E - E_\text{F}$ for the $\pi$ bands ($I_{\pi}$, red) and the background signal ($I_B$, blue) at $h\nu = \SIlist[list-units=single]{94; 100; 106}{\electronvolt}$. Intensities have been normalized to show average counts per pixel of the detector.}
\label{fig:Subtraction}
\end{figure}

The intensity curves in Fig.~\ref{fig:Subtraction} demonstrate the efficacy of the masking procedure outlined in Appendix \ref{mask}. While a distinct background artifact ($I_B$) is present at $E-E_\text{F} \approx \SI{-1.4}{{\electronvolt}}$ for $h\nu = \SI{100}{{\electronvolt}}$ [Fig.~\ref{fig:Subtraction}(b)], the extracted $\pi$ band intensity profile ($I_\pi$) preserves its line shape across all measured photon energies [Fig.~\ref{fig:Subtraction}(a)–(c)]. This confirms that the masking method successfully removes the Si $2p$ background artifact and that the peak in Fig.~\ref{fig:hvDependence}(c) and (d) at $h\nu = \SI{100}{\electronvolt}$ can be attributed to interlayer scattering and not just an artifact of the second harmonic leakage through the monochromator.

\appsection{Constructive Interference Condition for Interlayer Scattering} \label{derivation}

\begin{figure}[htbp]
\centering
\includegraphics[width=1\linewidth]{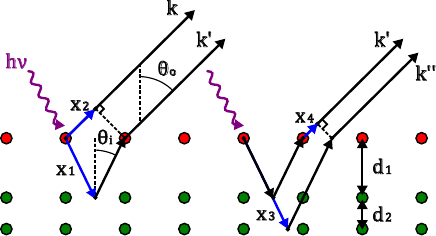}
\caption{Diagram of photoelectron scattering paths. The path $k$ represents direct emission from the surface layer, shown in red, to the detector while $k’$ and $k''$ represent the components scattered by lattice planes separated by a distance $d_1$ and $d_2$, shown in green. The wavevectors $k'$ and $k''$ refract at the surface due to the change in inner potential $V_0$ at the material/vacuum interface as illustrated by the internal and external emission angles $\theta_i$ and $\theta_o$. The lengths $x_i$ indicate different components of the paths traveled by the electron. }
\label{fig:App3}
\end{figure}

In photoemission, by changing the angle of a sample relative to the detector one can select the in-plane momentum $k_\parallel$ of the electrons that are measured. The angle $\theta_{o}$ required to select a particular location in momentum space can be calculated within the free electron model in terms of the kinetic energy $E$, mass $m_e$ of the electron, and the magnitude of $k_{ \parallel }$,
\begin{equation}
\theta_{o} = \sin^{-1} \left( \frac{\hbar k_{\parallel}}{\sqrt{2m_{e}E}} \right).
\end{equation}
\noindent In a similar manner, the angle inside the material ($\theta_i$) can be defined by conservation of energy and $k_{ \parallel }$ and the inner potential $V_0$ of the material:
\begin{equation}
\theta_{i} = \sin^{-1} \left( \frac{\hbar k_{\parallel}}{\sqrt{2m_{e}(E+V_0)}} \right).
\end{equation}
\noindent Because the crystal surface potential is periodic in the plane, parallel momentum is conserved across the surface interface ($k_{\parallel, i} = k_{\parallel, o} = k_\parallel$). Expressing these parallel components in terms of the total wavevector magnitudes inside ($k_i$) and outside ($k_o$) the material gives
\begin{equation}
k_i \sin\theta_i = k_o \sin\theta_o.
\end{equation}

In Fig. \ref{fig:App3}, a schematic of interlayer scattering of photoelectrons is shown. The path $k$ represents direct emission of the photoelectron from the material to the detector at an angle $\theta_o$. On the other hand, $k’$ represents a path where the electron diffracts from a lattice plane before traveling to the detector. For both paths to represent the same initial state, $\theta_o$ and the final kinetic energy $E$ must be the same. Since energy is conserved in this interaction, we know that the electron inside the material must have the same angle $\theta_i$ before and after diffracting from the lattice plane. This restricts the electron to a single path inside the material where the distance traveled from the emitting layer to the scattering plane is $x_1$. The distance $x_1$ can be shown to be $x_{1} = {d_1}/ \cos \theta_{i}$, yielding a total path length inside the material of $2x_1$. The segment $x_{2} = 2{d_1} \sin \theta_{o} \tan \theta_{i}$ is the projection of the lateral offset between the two surface exit points onto the direct wavevector $k$ to construct a plane of constant phase.

With the path lengths defined, we now consider the phase accumulation for both the direct and scattered paths. The wave vectors for these paths inside the material and in the vacuum are given by $k_i = \frac{\sqrt{2m_{e}(E + V_0)}}{\hbar}$ and $k_o = \frac{\sqrt{2m_{e}E}}{\hbar}$. The phase difference $\Delta \phi$ between the two paths is then determined by the difference in path lengths times the corresponding wave vectors. The total phase difference is expressed as
\begin{align}
\Delta\phi &= 2x_1 k_i - x_2 k_o \nonumber \\
&= 2\left(\frac{d_1}{\cos\theta_i}\right) k_i - \left(2d_1 \tan\theta_i \sin\theta_o\right) k_o \nonumber \\
&= \frac{2d_1 k_i}{\cos\theta_i} - 2d_1 \tan\theta_i (k_o \sin\theta_o) \nonumber \\
&= \frac{2d_1 k_i}{\cos\theta_i} - 2d_1 k_i \frac{\sin^2\theta_i}{\cos\theta_i} \nonumber \\
&= \frac{2d_1 k_i}{\cos\theta_i}\left(1 - \sin^2\theta_i\right) \nonumber \\
&= 2d_1 k_i \cos\theta_i,
\end{align}
\noindent and the final constructive interference condition for a general layer separation $d$ is
\begin{equation}
2\pi n = 2d k_i \cos\theta_i. \label{eq:interlayer_phase}
\end{equation}

This process can be repeated for the paths $k'$ and $k''$ to determine the constructive interference condition for electrons that scatter from two different lattice planes instead of one taking a direct path to the detector. In this case the phase difference can be expressed as in Eq.~\ref{eq:interlayer_phase} and depends on the separation $d_2$ between the planes.

It is instructive to compare this result with the photoemission diffraction framework of Fadley \cite{Fadley1984angle}. In the single scattering cluster approximation, a photoelectron emitted from a substrate atom and scattered by a neighboring atom at distance $r_j$ at an angle $\theta_j$ accumulates an overall path length difference between the direct and scattered paths of $\Delta L = r_j (1 - \cos\theta_j)$. The total phase shift is 
\begin{equation*}
\Delta\phi = k r_j (1- \cos\theta_j) + \psi_j(\theta_j),
\end{equation*}
\noindent where $\psi_j(\theta_j)$ is the phase shift associated with the atomic scattering event \cite{Fadley1984angle}.

In our model of planar interlayer scattering, the distance to the scatterer is $r_j = d / \cos\theta_i$ and the scattering angle is $\theta_j = \pi - 2\theta_i$. Substituting these parameters into the path difference term and setting $k=k_i$ yields
\begin{align*}
k_i r_j (1 - \cos\theta_j) &= k_i \left( \frac{d}{\cos\theta_i} \right) \left[ 1 - \cos(\pi - 2\theta_i) \right] \nonumber \\
&= k_i \left( \frac{d}{\cos\theta_i} \right) \left( 1 + \cos 2\theta_i \right) \nonumber \\
&= k_i \left( \frac{d}{\cos\theta_i} \right) \left( 2\cos^2\theta_i \right) \nonumber \\
&= 2d k_i \cos\theta_i.
\end{align*}
\noindent Thus, our simple geometric model recovers the exact spatial phase term $2d k_i \cos\theta_i$ of the full scattering theory, differing only by the atomic scattering phase shift $\psi_j(\theta_j)$. Because $\psi_j(\theta_j)$ scales with the atomic number $Z$ of the scattering species, we approximate $\psi_j(\theta_j) \approx 0$ given the light elements composing our sample ($Z_{\text{C}} = 6$, $Z_{\text{Si}} = 14$).

\appsection{Polarization Dependence of Intensity Resonances} \label{polarization}

\begin{figure}[htbp]
\centering
\includegraphics[width=1\linewidth]{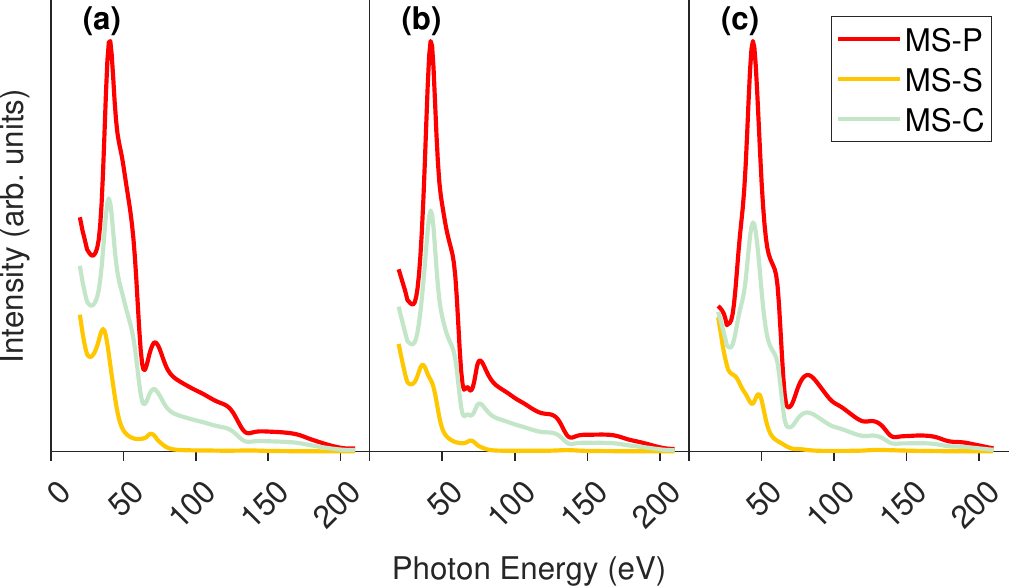}
\caption{Photon-energy-dependent photoemission intensity from MS simulations at binding energies of (a) $-0.25$ eV, (b) $-1.25$ eV, and (c) $-2.25$ eV. Each panel compares $p$-polarized (MS-P), $s$-polarized (MS-S), and circularly polarized (MS-C) light. Intensities are normalized to the maximum $p$-polarized intensity at each binding energy.}
\label{fig:pol}
\end{figure}

It is well established that photoemission intensity in ARPES depends on the dipole selection rules governing the photoemission matrix element, which can strongly influence the measured spectral intensity \cite{moser_experimentalists_2017}. To investigate the influence of light polarization on the photon-energy-dependent intensity resonances, MS simulations were performed using $p$-, $s$-, and circularly polarized light. Figure~\ref{fig:pol} compares the resulting photon-energy-dependent intensities for the three polarization configurations at three binding energies: \SIlist[list-units=single]{-0.25; -1.25; -2.25}{\electronvolt}. For each binding energy, the intensities are normalized to the maximum $p$-polarized intensity to facilitate comparison of the relative intensities and resonance positions. Although circular dichroism has been observed in materials such as graphite \cite{kruger2022observation}, in our experimental geometry, the opposite dichroic contributions from the two mirror-symmetry related branches ($k_x$ and $-k_x$) of the Dirac cone cancel when their intensities are summed.

The photon-energy-dependent intensity resonances observed with $p$-polarized light [Fig.~\ref{fig:hvDependence}(b)-(d)] are also present with both $s$- and circularly polarized light. However, their relative intensities and resonance positions exhibit some dependence on polarization. The $s$-polarized intensities are substantially lower than the $p$-polarized intensities. Additionally, the resonance locations generally shift toward lower photon energies for $s$-polarized light. For circularly polarized light, the resulting intensity is dominated by the much larger $p$-polarized contribution, producing a resonance profile that closely resembles that obtained with $p$-polarized light. These results demonstrate that, although the photoemission matrix elements influence the relative intensities and positions of the resonances, the photon-energy-dependent intensity modulations persist for all three polarization configurations.

%\bibliography{Graphene}


%\clearpage

\end{document}